\documentclass[twocolumn,           
               showpacs,            
               preprintnumbers,     
               aps,                 
               prd,          	    
               a4paper,             
               unsortedaddress,     
               nofootinbib,         
               tightenlines,        
               floats               
               ]{revtex4}

\usepackage{graphicx}
\usepackage{latexsym}
\usepackage{amsmath,amssymb}        
\usepackage[draft=false]{hyperref}

\newcommand{\de}{\mathrm{d}}
\renewcommand{\(}{\left(}
\renewcommand{\)}{\right)}

\newcommand{\period}{\,\mathrm{.}}
\newcommand{\comma}{\,\mathrm{,}}
\newcommand{\reffig}[1]{Fig.~\ref{#1}}

\newcommand{\refeqs}[2]{Eqs.~(\ref{#1})--(\ref{#2})}
\newcommand{\reftab}[1]{Table~\ref{#1}}

\newcommand{\mpl}{m_\mathrm{Pl}}

\renewcommand{\Re}{\mathrm{Re}}

\newcommand{\wf}{w_\mathrm{f}}
\newcommand{\wq}{w_Q}
\newcommand{\gf}{\gamma_\mathrm{f}}
\newcommand{\gq}{\gamma_Q}
\newcommand{\gtot}{\gamma_\mathrm{tot}}
\newcommand{\of}{\Omega_\mathrm{f}}
\newcommand{\oq}{\Omega_Q}
\newcommand{\rhf}{\rho_\mathrm{f}}
\newcommand{\rhq}{\rho_Q}
\newcommand{\csq}{c_{\mathrm{s}Q}^2}
\newcommand{\thq}{\theta_Q}
\newcommand{\drf}{\delta\rho_\mathrm{f}}
\newcommand{\drq}{\delta\rho_Q}
\newcommand{\dpf}{\delta p_\mathrm{f}}
\newcommand{\dpq}{\delta p_Q}
\newcommand{\dQ}{\delta Q}
\newcommand{\df}{\delta_\mathrm{f}}
\newcommand{\deq}{\delta_Q}
\newcommand{\dep}{\delta_P}
\newcommand{\xx}{\mathbf{x}}
\newcommand{\yy}{\mathbf{y}}
\newcommand{\matf}{\mathcal{F}}
\newcommand{\nua}{n_\mathrm{ut1}}
\newcommand{\nub}{n_\mathrm{ut2}}
\newcommand{\nuc}{n_\mathrm{ut3}}
\newcommand{\npa}{n_\mathrm{pt1}}
\newcommand{\npb}{n_\mathrm{pt2}}
\newcommand{\zz}{\mathbf{z}}
\newcommand{\matg}{\mathcal{G}}

\begin{document}


\title{Evolution of large-scale perturbations in quintessence models}
\author{Micha\"el Malquarti and Andrew R.~Liddle}
\affiliation{Astronomy Centre, University of Sussex, 
             Brighton BN1 9QJ, United Kingdom}
\date{\today} 
\pacs{98.80.Cq \hfill astro-ph/0208562}
\preprint{astro-ph/0208562}


\begin{abstract}
We carry out a comprehensive study of the dynamics of large-scale perturbations 
in quintessence scenarios. We model the contents of the Universe by a perfect 
fluid with equation of state $\wf$ and a scalar field $Q$ with potential $V(Q)$. 
We are able to reduce the perturbation equations to a system of four first-order 
equations. During each of the five main regimes of quintessence field behaviour, 
these equations have constant coefficients, enabling analytic solution of the 
perturbation evolution by eigenvector decomposition. We determine these 
solutions and 
discuss their main properties.
\end{abstract}

\maketitle

\section{Introduction}

Recent observations seem to indicate that the Universe is undergoing a period of 
accelerated expansion~\cite{acc}. Whereas cosmologists initially introduced a 
cosmological constant in order to explain this, a range of different models 
have emerged, amongst which quintessence has been particularly prominent in the 
literature~\cite{RP,qui}. It is defined as a scalar field rolling down its 
potential and presently dominating the dynamics of the Universe. An important 
class of quintessence models are known as tracking models~\cite{RP,qui}, where 
the 
late-time evolution of the field has an attractor behaviour rendering its 
evolution fairly 
independent of initial conditions. In contrast to a cosmological constant, which 
is by definition perfectly homogeneous, the quintessence field can, and indeed 
must, have perturbations.

The evolution of perturbations in quintessence models have been studied by many
authors \cite{RP,quintperts,BMR,perts2}.  In this paper we carry out an 
exhaustive
and elegant analysis of those in the large-scale approximation.  We model the
contents of the Universe by a perfect fluid with equation of state $\wf$ and a
scalar field $Q$ with potential $V(Q)$.  We assume a flat Universe throughout.

\section{Background evolution}

Before studying the perturbations, we recall some results for the homogeneous 
background evolution. The geometry of the Universe is described by a flat 
Robertson--Walker metric
\begin{equation}
\de s^2= - \de t^2 + a^2(t)\de\mathbf{x}^2\period
\end{equation}
The Einstein equations
\begin{eqnarray}
H^2\,=\,\Big(\frac{\dot{a}}{a}\Big)^2
	&=&\frac{8\pi}{3\mpl^2}\(\rhf+\rhq\)\,,\\
2\dot{H}+3H^2\,=\,2\frac{\ddot{a}}{a}+\Big(\frac{\dot{a}}{a}\Big)^2
	&=&-\frac{8\pi}{\mpl^2}\(p_\mathrm{f}+p_Q\) 
\end{eqnarray}
relate the matter components to the geometry. The indices ``f'' and ``$Q$'' 
always refer to the perfect fluid and the quintessence field respectively, and 
dots are time derivatives. We 
will use a prime to denote a derivative with respect to $N\equiv\log(a/a_0)$. 

The evolution of the fluid is straightforward, with its energy density scaling 
as $a^{-3(1+\wf)}$, where $\wf$ is the ratio of pressure to energy density of 
the fluid. The quintessence field follows the Euler--Lagrange equation
\begin{equation}
\ddot{Q}=-3H\dot{Q}-\frac{\de V}{\de Q}
\end{equation}
and its equation of state is
\begin{equation}
\wq\equiv\frac{p_Q}{\rho_Q}=\frac{\dot{Q}^2/2-V(Q)}{\dot{Q}^2/2+V(Q)} 
\period
\end{equation}

Depending on the precise model and on the choice of initial conditions, the 
quintessence dynamics can feature up to five main regimes, which were classified 
by Brax et al.~\cite{BMR} and which appear in a sequential order. During the 
first three the quintessence field is sub-dominant.\footnote{Actually, one can 
consider initial conditions with domination of the quintessence field, but they 
usually lead to an ``overshoot'' of the required present energy density, and 
are therefore not interesting.} The ``kinetic'' regime is characterized by the 
domination of the kinetic energy which scales as $a^{-6}$. In the ``transition'' 
and ``potential'' regimes the potential energy dominates and the energy density 
remains constant. The sound speed of the quintessence field is defined by  
\begin{equation}
\label{csdef}
\csq\equiv\frac{\dot{p}_Q}{\dot{\rho}_Q}
=\wq-\frac{\wq'}{3(1+\wq)}
=1+\frac{2}{3}\frac{\de V/\de Q}{H\dot{Q}}\comma
\end{equation} 
and is equal to $1$ or $-2-\wf$ respectively during those regimes. During the 
``tracker'' regime the quintessence field approximately mimics the behaviour of 
the fluid, and usually its energy density is still sub-dominant when tracking 
begins. If $\wq=\wf$ there is perfect tracking, even if the 
quintessence field is not sub-dominant. Finally the field enters its 
``domination'' phase, during which $\wq$ tends to $-1$ in most cases.

\begin{figure}[t]
\includegraphics[scale=0.38,angle=-90]{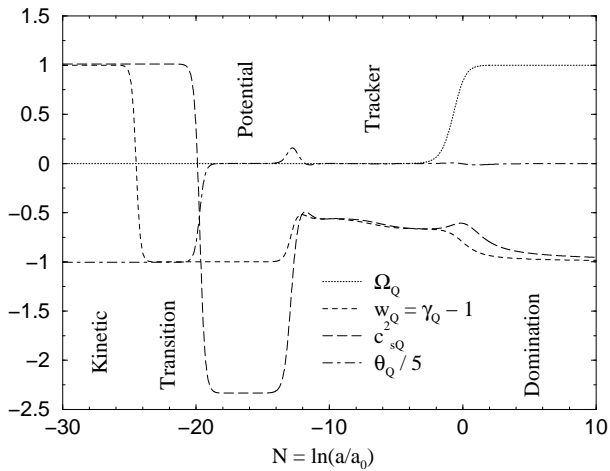}
\caption{An example of the evolution of the four parameters $\oq$, $\wq$, 
$\csq$, and $\thq/5$ in a realistic Universe, showing the five different regimes 
and also the transition between radiation domination and matter domination (at 
$N\simeq-7$). We used an inverse power-law potential 
$V(Q)=V_0(Q/\mpl)^{-\alpha}$, with $\alpha=1$ and $V_0=3\times 
10^{-124}\,\mpl^4$. In this case we have the usual sub-dominant tracker regime.}
\label{regimes}
\end{figure}

Four useful parameters can be defined to describe these five regimes. They are 
the quintessence density parameter $\oq$, the equation of state conveniently 
parametrized as $\gq\equiv 1+\wq$, the speed of sound $\csq$ as defined in 
Eq.~(\ref{csdef}), and one further 
parameter relating to the speed of sound defined as 
\begin{equation}
\thq\equiv\frac{\left(\csq\right)^\prime}{1-\csq}
=-3(1+\csq)-\frac{\de}{\de N} \log\(Q' \de V/\de Q \)\period
\end{equation}
In order to simplify the notation, we define the vector 
$\xx\equiv (\oq,\gq,\csq,\thq)$. In \reffig{regimes}, we show an example of the 
evolution of those parameters. We take the realistic case $\wf=1/3$ during 
radiation domination and $\wf=0$ during matter domination, and we use an inverse 
power-law quintessence potential. We clearly see the five different regimes, and 
also the transition between radiation domination and matter domination which in 
this case occurs during the tracking regime. In \reftab{val_reg} we give the 
values of the parameters in the general case for each regime.

\vspace*{4cm}  

\begin{table}[t]
\begin{tabular}{l|cccc}
\hline
\hline
                      &
\makebox[1cm]{$\oq$}  & 
\makebox[1cm]{$\gq$}  &
\makebox[1cm]{$\csq$} &
\makebox[1cm]{$\thq$} \\
\hline
Kinetic               & 0     & 2     & 1        & $-3(3+\wf)/2$ \\
Transition            & 0     & 0     & 1        & $-3(3+\wf)/2$ \\
Potential             & 0     & 0     & $-2-\wf$ & 0             \\
Usual Tracker         & 0     & $\gq$ & $\wq$    & 0             \\
Perfect Tracker       & $\oq$ & $\gf$ & $\wq$    & 0             \\
Domination            & 1     & $\gq$ & $\wq$    & 0             \\
\hline
\hline
\end{tabular}
\caption{Values of the four parameters $\oq$, $\gq$, $\csq$, and $\thq$ 
during the different regimes.}
\label{val_reg}
\end{table}

\section{Perturbation evolution}

In order to describe the perturbations we choose the conformal Newtonian 
gauge~\cite{MFB}. As long as there is no anisotropic stress, the perturbed 
metric is
\begin{equation}
\de s^2=-(1+2\Phi)\de t^2+a^2(t)(1-2\Phi)\de \mathbf{x}^2\comma
\end{equation}
where in this case $\Phi$ is equal to the gauge-invariant potential defined as 
in Ref.~\cite{MFB}. We 
work in Fourier space and compute the first-order perturbed Einstein 
equations
\begin{eqnarray}
-3H(H\Phi+\dot{\Phi})-\frac{k^2}{a^2}\Phi&=&4\pi G(\drf+\drq)\comma\\
\ddot{\Phi}+4H\dot{\Phi}+(2\dot{H}+3H^2)\Phi&=&4\pi G(\dpf+\dpq)\comma
\end{eqnarray}
where the perturbed fluid pressure $\dpf=\wf\drf$ and the perturbations in the 
quintessence energy density and pressure are given by
\begin{eqnarray}
\drq&=&\dot{Q}\delta\dot{Q}-\dot{Q}^2\Phi+\frac{\de V}{\de Q} \dQ\comma\\
\dpq&=&\dot{Q}\delta\dot{Q}-\dot{Q}^2\Phi-\frac{\de V}{\de Q} \dQ\period
\end{eqnarray}
The perturbed Euler--Lagrange equation and fluid conservation equation lead to
\begin{align}
&\delta\ddot{Q}+3H\delta\dot{Q}+\frac{k^2}{a^2}\dQ+\frac{\de^2 V}{\de Q^2}\dQ
=4\dot{Q}\dot{\Phi}-2\frac{\de V}{\de Q}\Phi\comma\\
&\dot{\df}-3(1+\wf)\dot{\Phi}=-(1+\wf)\frac{k}{a}\mathcal{V}_\mathrm{f}\comma
\end{align}
where $\df\equiv\drf/\rhf$ and $\mathcal{V}_\mathrm{f}$ gives the fluid 
velocity. 
{}From now on we use $N$ as a time variable and study the evolution of $\Phi$, 
$\df$, $\deq\equiv\drq/\rhq$ and $\dep\equiv\dpq/\rhq$ in the long-wavelength 
limit ($k/aH\ll1$). We define the vector $\yy\equiv(\Phi,\df,\deq,\dep)^T$, and 
considerable algebra leads to the expression
\begin{equation}\label{lin_sys}
\yy'=\matf(\gf,\xx)\times\yy\comma
\end{equation}
where the matrix $\matf(\gf,\xx)$ is given by
\begin{widetext}
\begin{equation}
\label{Fmat}
\matf(\gf,\xx)=\(\begin{array}{cccc}
 -1        & -\of/2     & -\oq/2          & 0 \\
 -3\gf     & -3\gf\of/2 & -3\gf\oq/2      & 0 \\ 
 -3\gq     & -3\gq\of/2 & -3\gq\oq/2+3\wq & -3 \\
 -3\gq\csq & -3\gq\of/2 & 3\gf\of/2+\thq  & 
 -3\gq\oq/2+3\wq-3\gf\of/2-\thq-3\csq  
\end{array}\)\period
\end{equation}
\end{widetext}
In order to display the above expression in compact form we used the variables 
$\of=1-\oq$ and 
\mbox{$\wq=\gq-1$}, but the matrix depends only on the five independent 
parameters 
$\gf$, $\oq$, $\gq$, $\csq$ and $\thq$. Recall that $\gf$ and $\gq$ take on 
values between $0$ and $2$. 

The general evolution is of course very complicated, but as seen in 
\reffig{regimes}, during each of the five regimes described above the 
coefficients within the matrix take on constant values, and this allows us to 
study the main features of the perturbation evolution. We are interested in the 
eigenvalues $n_i$ of the matrix $\matf$ and their corresponding eigenvectors 
${\bf y}_i$. The solution 
then takes the form 
\begin{equation}
{\bf y} = \sum_{i=1}^4 A_i \, {\bf y}_i \exp(n_iN) \,,
\end{equation}
where the $A_i$ are constants given by the initial conditions.

\subsection{The adiabatic case}

Before considering the general case, we restrict ourselves to adiabatic 
perturbations. For perturbations to be adiabatic, they must share a common 
perturbation according to the prescription
\begin{equation}
\frac{\drf}{\dot{\rho}_\mathrm{f}} =
\frac{\drq}{\dot{\rho}_Q} =
\frac{\dpq}{\dot{p}_Q} \comma
\end{equation}
which ensures that all matter perturbations vanish on uniform-density 
hypersurfaces. Note that the quintessence pressure perturbation, as well as its 
density perturbation, must satisfy the adiabatic condition (for the perfect 
fluid adiabaticity of its pressure perturbation is automatically guaranteed by 
its equation of state).
These conditions can be rewritten as
\begin{equation}\label{adiabatic}
\frac{\df}{\gf} = \frac{\deq}{\gq} = \frac{\dep}{\csq\gq} \period
\end{equation}

It is well known that initially adiabatic perturbations remain purely adiabatic, 
and indeed it is not difficult to check that these conditions are conserved 
through 
evolution by our equations. We can therefore reduce the dynamical system in the 
adiabatic case to a 
system of two 
first-order equations. We define the vector $\zz\equiv(\Phi,\df)^T$, and using 
Eqs.~(\ref{lin_sys}) and (\ref{adiabatic}) we find
\begin{equation}
\zz'=\matg(\gf,\oq,\gq)\times\zz\comma
\end{equation}
where the matrix $\matg(\gf,\oq,\gq)$ is given by
\begin{equation}
\matg(\gf,\oq,\gq)=\(\begin{array}{cc}
 -1        & -\gtot/2\gf \\
 -3\gf     & -3\gtot/2 
\end{array}\)\comma
\end{equation}
where $\gtot=\gf\of+\gq\oq$. The eigenvalues and eigenvectors are given in the 
upper part of \reftab{results}. We see, as is well known, that there are always 
a constant and a decaying adiabatic mode, the former giving the late-time 
solution $\Phi = -\gtot \df/2\gf = -\delta_{{\rm tot}}/2$, where $\delta_{{\rm 
tot}}\equiv \delta\rho_{{\rm tot}}/\rho_{{\rm tot}}$. 

\begin{table}[t!]
\begin{tabular}{lcl}
\hline
\hline
&&\\
\multicolumn{3}{l}{Adiabatic perturbations}\\
$n_1=0$           && $\zz_1=(-\gtot/2,\gf)$ \\
$n_2=-1-3\gtot/2$ && $\zz_2=(1/3,\gf)$\\
&&\\
\hline
\hline
&&\\
\multicolumn{3}{l}{Kinetic regime}\\
$n_1=0$          && $\yy_1=(-\gf/2,\gf,2,2)$ \\
$n_2=-1-3\gf/2$  && $\yy_2=(1/3,\gf,2,2)$ \\
$n_3=+6$         && $\yy_3=(0,0,1,-1)$\\
$n_4=0$          && $\yy_4=(0,0,1,1)$ \\
&&\\

\multicolumn{3}{l}{Transition regime}\\
$n_1=0$          && $\yy_1=(-1/2,1,0,0)$\\
$n_2=-1-3\gf/2$  && $\yy_2=(1/3,\gf,0,0)$\\
$n_3=0$          && $\yy_3=(0,0,1,-1)$\\
$n_4=-6$         && $\yy_4=(0,0,1,1)$\\
&&\\

\multicolumn{3}{l}{Potential regime}\\
$n_1=0$          && $\yy_1=(-1/2,1,0,0)$\\
$n_2=-1-3\gf/2$  && $\yy_2=(1/3,\gf,0,0)$\\
$n_3=0$          && $\yy_3=(0,0,1,-1)$\\
$n_4=-3+3\gf/2$  && $\yy_4=(0,0,-2,\gf)$\\
&&\\

\multicolumn{3}{l}{Usual tracker regime}\\
$n_1=0$          && $\yy_1=(-\gf/2,\gf,\gq,\wq\gq)$\\
$n_2=-1-3\gf/2$  && $\yy_2=(1/3,\gf,\gq,\wq\gq)$\\
$n_3=\nua+\nuc$  && $\yy_3=(0,0,\nub+\nuc,3\gf/2)$\\ 
$n_4=\nua-\nuc$  && $\yy_4=(0,0,\nub-\nuc,3\gf/2)$\\
&&\\

\multicolumn{3}{l}{Perfect tracker regime}\\
$n_1=0$          && $\yy_1=(-\gf/2,\gf,\gf,\wf\gf)$\\
$n_2=-1-3\gf/2$  && $\yy_2=(1/3,\gf,\gf,\wf\gf)$\\
$n_3=\npa+\npb$  && $\yy_3=\ldots$\\ 
$n_4=\npa-\npb$  && $\yy_4=\ldots$\\
&&\\

\multicolumn{3}{l}{Domination regime}\\
$n_1=0$          && $\yy_1=(-\gq/2,\gf,\gq,\wq\gq)$\\
$n_2=-1-3\gq/2$  && $\yy_2=(1/3,\gf,\gq,\wq\gq)$\\ 
$n_3=0$          && $\yy_3=(0,1,0,0)$\\
$n_4=-3+3\gq/2$  && $\yy_4=(1/3,\gf,1/3-\wq,-\gq/3)$\\
&&\\
\hline
\hline
\end{tabular}
\caption{Top: eigenvalues and eigenvectors of the matrix 
$\matg$ (adiabatic case). Bottom: eigenvalues and eigenvectors of the matrix 
$\matf$ according to 
the different regimes; $\nua$, $\nub$, $\nuc$, $\npa$ and $\npb$ are given in 
\refeqs{nua}{npb}.}
\label{results}
\end{table}

\subsection{The general case}

We now return to the full set of perturbation equations Eq.~(\ref{lin_sys}), 
continuing to consider the regimes in each of which the coefficients of the 
matrix $\matf$ remain constant.
We summarize our main results in the lower part of \reftab{results}. For the 
perfect tracker 
regime the eigenvectors $\yy_3$ and $\yy_4$ have not been given, since they are 
long and complicated formulas which are anyway not very relevant. 
In order to simplify some expressions we have used the variables
\begin{eqnarray}
\nua & = & -\frac{3}{4}(\gf-2\wq)\comma \label{nua}\\
\nub & = & \frac{3}{4}(\gf+2\wq)\comma\\
\nuc & = & \frac{3}{4}\sqrt{(2\wq+\gf)^2-8\gf}\comma\\
\npa & = & -\frac{3}{4}\(2-\gf\)\comma\\
\npb & = & \frac{3}{4}\sqrt{(2-\gf)(2-\gf-8\gf\of)}\period \label{npb}
\end{eqnarray}

For each regime the adiabatic modes can easily be identified as the first two 
entries in the table. Now let us analyze 
the two other, non-adiabatic, modes. In the 
kinetic case there are a growing mode, for which $\dep=-\deq$ and thus 
$\delta\dot{Q}=0$ (since $\Phi = 0$), and another constant mode corresponding to 
$\delta Q = 0$. During the transition and 
potential regimes the former growing mode becomes constant and for each regime 
the fourth mode is decaying. Therefore, before entering the tracker the 
quintessence field may feature large non-adiabatic perturbations. As long as the
Universe is dominated by the fluid, they are isocurvature perturbations.

\begin{figure}[t]
\includegraphics[scale=0.38,angle=-90]{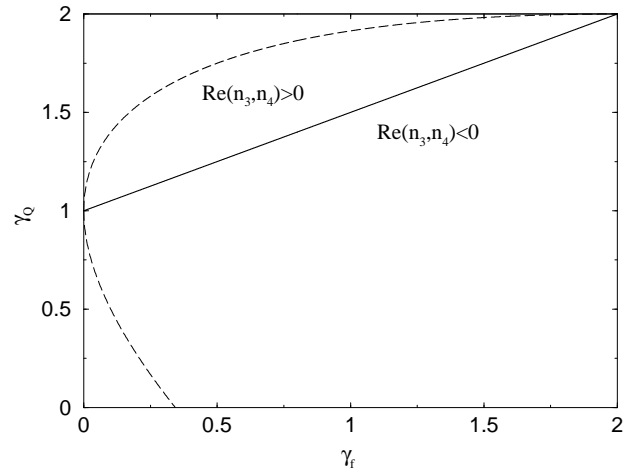}
\caption{Features of the eigenvalues $n_3$ and $n_4$ in the case of an usual 
tracker regime, as a function of $\gf$ and $\gq$. Their real part is negative 
below the solid line and they have an imaginary part on the right-hand side of 
the 
dashed line.}
\label{eigenvalues}
\end{figure}

In the usual tracker case the last two eigenvalues may have an imaginary part, 
leading to oscillations, and their real part can either be negative or 
positive according to the value of $\gf$ and $\gq$. The regimes of the 
eigenvalues are shown in \reffig{eigenvalues}.
However, since the 
quintessence field has to dominate at the present epoch, $\rhq$ must decrease 
more slowly than $\rhf$, and hence $\gq<\gf$. As one can see in 
\reffig{eigenvalues}, 
this implies $\Re(n_{3,4})<0$ . In the perfect tracker case one easily sees 
that the last two modes decay, possibly oscillating as they do. As 
long as all the other modes are decaying, during the tracker regime the constant 
adiabatic mode $\yy_1$ is an attractor. As a result, a long tracker period 
implies the suppression of all non-adiabatic modes \cite{BMR,perts2}. Moreover, 
in the case of a 
sub-dominant tracker the late-time evolution of the perturbations is even 
independent of the quintessence field initial conditions.

Finally, during the domination regime, which is reached after the present 
epoch, the non-adiabatic modes are constant and decaying.

\section{Conclusions}

We have derived four first-order equations to describe large-scale 
perturbations in quintessence scenarios. During each of the five main 
regimes of quintessence behaviour, these equations have constant coefficients, 
enabling analytic solution of the perturbations by eigenvector decomposition.

We have seen that during the kinetic regime there is a growing isocurvature mode 
which then remains constant until tracking begins. However, if the quintessence 
field undergoes a long period of tracking, there remain only adiabatic 
perturbations which, in case of a sub-dominant tracker, are independent of its 
initial conditions. A low initial quintessence energy density~\cite{ML}, or a 
long kinetic period, may prevent the non-adiabatic modes disappearing completely.

It is possible in principle to carry out the same analysis without using the 
large-scale approximation, although the equations may then be too 
complicated 
to be useful.


\begin{acknowledgments}
M.M.~was supported by the Fondation Barbour, the Fondation Wilsdorf and the 
Janggen-P\"{o}hn-Stiftung, and A.R.L.~in part by the Leverhulme Trust. We thank 
Pier-Stefano Corasaniti for useful discussions.
\end{acknowledgments}



\begin{thebibliography}{}
\bibitem{acc} A. G. Riess et al., Astron. J. \textbf{116}, 1009 (1998),
        \texttt{astro-ph/9805201}; P. Garnavich et al., Ap. J. \textbf{509}, 74
        (1998), \texttt{astro-ph/9806396}; S. Perlmutter et al., Ap. J.
        \textbf{517}, 565 (1998), \texttt{astro-ph/9812133}; G. Efstathiou 
        et al., Mon. Not. Roy. Ast. Soc. \textbf{330}, L29 (2002), 
        astro-ph/0109152.
\bibitem{RP} B. Ratra and P. J. E. Peebles, Phys. Rev. D\textbf{37}, 3406
        (1988).
\bibitem{qui} C. Wetterich, Nucl. Phys. B\textbf{302}, 668 (1988); 
        E. J. Copeland, A. R. Liddle, and D. Wands, Ann. N. Y. Acad. 
	Sci. {\bf 688},	647 (1993); R. R. Caldwell, R. Dave, and P. J. 
	Steinhardt, Phys. Rev. Lett. \textbf{80}, 1582 (1998),
	\texttt{astro-ph/9708069}; P. G. Ferreira and M. Joyce, Phys. Rev. 
	D\textbf{58}, 023503 (1998), \texttt{astro-ph/9711102}; E. J. 
	Copeland, A. R. Liddle, and D. Wands, Phys. Rev. 
	D{\bf 57}, 4686 (1998), {\tt gr-qc/9711068}; I. Zlatev, L. Wang, and 
	P. J. Steinhardt, Phys. Rev. Lett. \textbf{82}, 896 (1999),
        \texttt{astro-ph/9807002}; A. R. Liddle and R. J. Scherrer, Phys. 
        Rev. D{\bf 59}, 023509 (1999), {\tt astro-ph/9809272}; V. Sahni and
        A. Starobinsky, Int. J. Mod. Phys. \textbf{D9}, 373 (2000), 
        \texttt{astro-ph/9904398}.
\bibitem{quintperts} R. R. Caldwell, R. Dave, and P. J. Steinhardt,
	Phys. Rev. Lett. {\bf 80}, 1582 (1998), \texttt{astro-ph/9708069};
	P. T. P. Viana and A. R. Liddle, Phys. Rev. D\textbf{57}, 674 (1998),
	\texttt{astro-ph/9708247}; F. Perrotta and C. Baccigalupi, Phys. Rev.
	D\textbf{59}, 123508 (1999), \texttt{astro-ph/9811156}. 
\bibitem{BMR} P. Brax, J. Martin, and A. Riazuelo, Phys. Rev. D\textbf{62},
	103505 (2000), \texttt{astro-ph/0005428}.
\bibitem{perts2} L. R. Abramo and F. Finelli, 
	Phys. Rev. D\textbf{64}, 083513 (2001), \texttt{astro-ph/0101014}; 
	M. Kawasaki, T. Moroi, and T. Takahashi, Phys. Lett. B {\bf
	533}, 294 (2002), \texttt{astro-ph/0108081};
	R. Dave, R. R. Caldwell, and Paul J. Steinhardt, Phys. Rev.
	D\textbf{66}, 023516 (2002), \texttt{astro-ph/0206372}.
\bibitem{MFB} V. Mukhanov, H. Feldman, and R. Brandenberger, Phys. Rep.
        \textbf{215}, 203 (1992).
\bibitem{ML} M. Malquarti and A. R. Liddle, Phys. Rev. D\textbf{66},
        023524 (2002), \texttt{astro-ph/0203232}.
\end{thebibliography}
\end{document}